\documentclass[amsmath,amssymb,prl,10pt,showpacs,twocolumn,superscriptaddress]{revtex4}
\usepackage{times}
\usepackage{textcomp}
\usepackage{amsmath}
\usepackage{graphicx} 
\usepackage{dcolumn}  
\usepackage{bm}       
\usepackage[english]{babel}
\begin{document}


\newcommand{\gtwo}[1]{\mbox{$\mathrm{g}^{(2)}(#1)$}}
\newcommand{\Gtwo}[1]{\mbox{$\mathrm{G}^{(2)}(#1)$}}
\newcommand{\lag}{\mbox{$\delta{}t$}}


%
\author{Dennis Heine}
\affiliation{%
Atominstitut der \"Osterreichischen Universit\"aten, Technische Universit\"at Wien, Stadionallee 2, 1020 Wien, Austria}%
\affiliation{%
Physikalisches Institut, Universit\"at Heidelberg, Philosophenweg 12, 69120 Heidelberg, Germany}%
\author{Marco Wilzbach}
\affiliation{%
Atominstitut der \"Osterreichischen Universit\"aten, Technische Universit\"at Wien, Stadionallee 2, 1020 Wien, Austria}%
\affiliation{%
Physikalisches Institut, Universit\"at Heidelberg, Philosophenweg 12, 69120 Heidelberg, Germany}%
\author{Thomas Raub}
\affiliation{%
Atominstitut der \"Osterreichischen Universit\"aten, Technische Universit\"at Wien, Stadionallee 2, 1020 Wien, Austria}%
\affiliation{%
Physikalisches Institut, Universit\"at Heidelberg, Philosophenweg 12, 69120 Heidelberg, Germany}%
\author{Bj\"orn Hessmo}
\email{hessmo@atomchip.org}
\affiliation{%
Atominstitut der \"Osterreichischen Universit\"aten, Technische Universit\"at Wien, Stadionallee 2, 1020 Wien, Austria}%
\affiliation{%
Physikalisches Institut, Universit\"at Heidelberg, Philosophenweg 12, 69120 Heidelberg, Germany}%
\author{J\"org Schmiedmayer}
\affiliation{%
Atominstitut der \"Osterreichischen Universit\"aten, Technische Universit\"at Wien, Stadionallee 2, 1020 Wien, Austria}%
\affiliation{%
Physikalisches Institut, Universit\"at Heidelberg, Philosophenweg 12, 69120 Heidelberg, Germany}%
\title{An integrated atom detector: single atoms and photon statistics}
\pacs{03.75.-b, 42.50.Lc, 42.81.-i, 07.60.Vg}
\begin{abstract}
We demonstrate a robust fiber optics based fluorescence detector, fully integrated on an atom chip, which detects single atoms propagating in a guide with 66\% efficiency. We characterize the detector performance and the atom flux by analysing the photon statistics. Near-perfect photon antibunching proves that single atoms are detected, and allows us to study the second-order intensity correlation function of the fluorescence over three orders of magnitude in atomic density.\end{abstract}

\date{\today}

\maketitle

The ability to efficiently detect single particles is of
fundamental importance to many branches of science. For example
counting single photons was one of the starting points of quantum
optics \cite{man1995} and single particle (qubit) detection is one
of the key ingredients for quantum technologies \cite{Bow00}.

Detecting atoms is usually achieved by illumination with
near-resonant light, followed by a measurement of the absorption,
phase shift or fluorescence. Placing an optical cavity around the
detection region significantly enhances the signal and single atom
sensitivity can be achieved
\cite{mab1996,mun1999,hoo2000,ott2005,Haa06,aok2006,tep2006,col2007,tru2007}.
Active alignment of the cavities is however technically
challenging.

Single pass absorption does not allow one to detect single free
atoms \cite{vEnk01}, is necessary to hold the atom in a trap to
reach sufficiently long integration times \cite{quintosu2004,tey2008}. Similarly,
fluorescence detection is very efficient if atoms are tightly
localized in a trap and many photons can be collected. It is the
method of choice for many single atom or ion experiments
\cite{kuh2001,row2001,lei2003,dar2005,vol2006,web2006,sor2007}.
 Free neutral atoms are considerably harder to detect because the few scattered photons are difficult to distinguish from background light. 
Recently, fluorescence detection of freely falling single atoms has been demonstrated by using
macroscopic mirrors covering a solid angle of almost $4\pi$ \cite{bon2006}.

In an ideal fluorescence detector, the background is negligible and a single detected photon implies that an atom is present in the detection region. 
In this letter we present a simple fluorescence detector based on fiber optics fully integrated on an atom chip \cite{Fol02,Zimmermann2007} that detects single atoms in a magnetic guide with high signal to noise ratio \endnote{Signal to noise ratio (SNR) is defined as the signal count rate divided by the count rate of the background. In our present experiments it ranges between 20 and $>$100 and is only limited by the dark counts of the APD.} and an efficiency of 66\%.  Previous implementations of on chip detectors have been based on absorption
detection where small atomic ensembles (typical 100 atoms) could
be detected \cite{quintosu2004}, or used cavities \cite{tep2006,
col2007,tru2007}, which are more complex to handle.

\begin{figure}[t!]
    \includegraphics[width=0.48\textwidth]{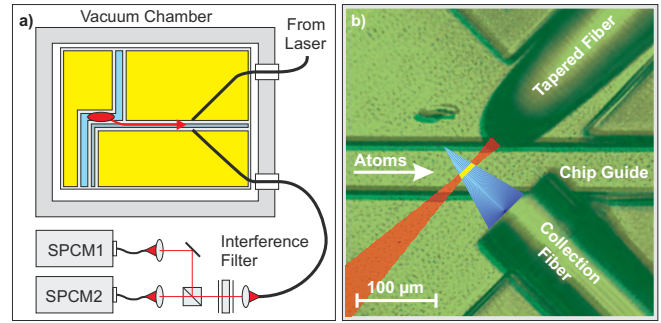}%
    \caption{(color online).
 a) Basic layout of the detector and the atom chip in the vacuum
    chamber. Atoms are initially trapped in a magnetic trap generated by a Z-shaped wire.
    A magnetic guide transports the atoms to the detector.
    The excitation light is delivered by tapered fiber,
    the fluorescence light is collected by a multimode fiber.
    Both are connected to the optics outside the vacuum chamber by
    a fiber feed-through \protect{\cite{abr1998}}.
    The collected light passes an interference filter (centered at 780 nm with 3 nm  bandwidth) 
    before detected by the single photon counting module(s) (SPCM).
    For high efficiency atom detection a single SPCM is employed while
    correlation measurements require two SPCMs in Hanbury Brown-Twiss like configuration.
 b) A microscopy image of the detection region on the chip.
    The multimode fiber collects light from a cone (blue) determined by its NA.
    The overlap of this cone with the excitation light from the tapered fiber (red)
    defines the detection region (yellow).}
    \label{fig:chippic}
\end{figure}

Our detector consists of a tapered lensed single mode fiber (focal
length of 40~\textmu m and a mode diameter of 5~\textmu m) used to
excite the atoms in a small detection region, the fluorescence
photons are selectively collected from this small volume by a
standard gradient-index multimode fiber with low mode dispersion,
and numerical aperture (NA) of 0.275, mounted at an angle
of 90 degrees (Fig.~\ref{fig:chippic}b) to reduce stray light
\endnote{The advantages of this setup are
related to properties of a confocal microscope: The tapered lensed fiber
delivers the excitation light efficiently to a small detection
region and the multimode fiber selectively collects the
fluorescent photons from this small volume and a large fraction
of the background noise is filtered away very efficiently, even
when the detection region is located only 62.5~\textmu m above
the chip surface. The difference to a regular confocal microscope
is that the involved point spread functions are different.}. The
collected light is sent through an interference filter to photon
counter(s) (SPCM) as illustrated in Fig.~\ref{fig:chippic}a. The
overall probability of detecting a fluorescent photon in this
setup is \mbox{$p_\mathrm{det} =0.9$\%}, including the collection
efficiency, losses, and the SPCM efficiency.

The detector is fully integrated on our atom chip
\cite{groth2004} by mounting the two fibers on the chip surface in
lithographically defined holders fabricated from SU-8 resist which
allow very stable passive alignment.
\cite{liu2005,wil2006}.

The fiber-based detector presented here has exceptionally low
background, the dominating contribution are the dark counts ($\sim
250$~cps) of the employed photon counter (Perkin-Elmer,
SPCM-AQR-12). Despite the proximity of the chip surface, the
influence of stray probe-light is essentially eliminated with a
suppression of better than $10^{-8}$. 1~nW excitation light
contributes only $\sim 30$~cps to the background. This extremely
low background allows high fidelity detection of a single atom and
accurate measurements of photon and atom statistics.

The experiment is carried out in a setup similar to the one
described in \cite{wil2004}. ${}^{87}$Rb atoms are first laser
cooled in a magneto-optic trap, optically pumped into the
\mbox{$|F=2,m_F=2\rangle$} state, and transferred to a
Ioffe-Pritchard type magnetic trap generated by a Z-shaped wire on
the chip surface. The atoms are then transferred into a magnetic
guide where the atomic cloud can expand towards the
detector situated $5.5$ mm away from the magnetic chip trap used
to prepare the atoms (Figure~\ref{fig:chippic}a). The phase space
density in the magnetic trap and guide is always less than
$10^{-5}$.

\begin{figure}[t!]
    \includegraphics[width=0.48\textwidth]{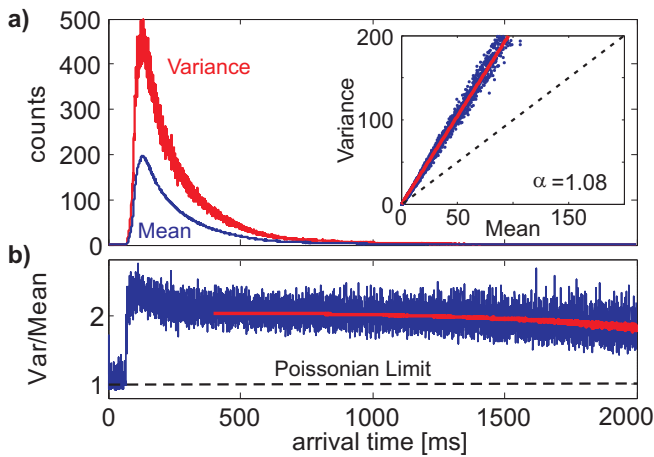}%
    \caption{(color online).
    Mean and variance of the photon counts.
    \emph{a)} Mean (blue) and variance (red) of the collected photons
    in 200 \textmu s time bins obtained from 600 experimental runs.
    Inset: Variance as a function of the mean to determine $\alpha $. A fit to the data according to
    Eq.~\protect{\ref{eq:poissonianAtomsAndBg}} gives $\alpha=1.08$
    counts per atom. The dashed line corresponds to Poissonian photon statistics.
   \emph{b)} Ratio variance over mean:  In the first 50~ms, where no atoms are present,
    the collected background follows Poissonian statistics.
    As soon as the atoms arrive var/mean exhibits a sudden jump to a
    superpoissonian value.
    The initial overshoot is an artifact created by significant change
    in count rate within a single bin.
    At long times the measured ratio decreases as random background
    counts become more important with decreasing atom flux.
    The red line gives a fit according to Eq.(\ref{eq:poissonianAtomsAndBg})
    over more than two orders of magnitude in atom flux.
    }
    \label{fig:varandmean}
\end{figure}

The position of the magnetic guide above the chip surface is
aligned with the focus of the tapered lensed fiber at the
detection region by adjusting the current through the chip wire
and the strength of the external magnetic field. Atoms passing
through the focus of the lensed fiber are excited by laser light
tuned near the $F=2$ to $F'=3$ transition in ${}^{87}$Rb.
Over the next 2000 ms atoms pass the detector and the arrival times of the fluorescent photons are registered. The experiment is repeated several times to measure the photon
statistics (Fig.~\ref{fig:varandmean}).

One observation from these measurements is that the
effects of stray light on the guided atoms can be neglected. This
is quite remarkable, because magnetic traps are extremely
sensitive to the presence of light close to resonance with an
atomic transition. On average, scattering of a little more than a
single photon is sufficient to pump the atom into a magnetically
untrapped state, removing it from the magnetic guide.

\begin{figure}[t!]
    \includegraphics[width=0.5\textwidth]{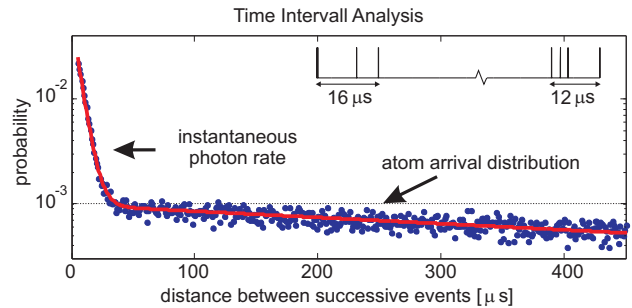}%
    \caption{(color online).
    Time interval analysis.
    The data points show the probabilities for finding different time intervals between
    successive photon detections in the low density tail of the atom distribution.
    The red line is a double exponential fit.
    Short time intervals are dominated by the fluorescence rate of the single
    atoms. Two such burst events are shown in the inset. The typical interaction 
    time of the atoms is $\tau{}=12$~\textmu s.
    The long intervals are given by the atom arrival rate at the detector.
    The good agreement with the exponential fit demonstrates that atoms arrive stochastically independent.
    The slope of the measured distribution is in agreement with the mean rate of atom detections.}
    \label{fig:TIA}
\end{figure}

When an atom arrives at the detector, it absorbs and then re-emits
photons. A few of these photons are counted by the SPCM. This
photon scattering strongly disturbs the atom, will pump it into a
un-trapped state, or even into a different hyper fine ground state
which is not excited by the probing light.  Consequently after a
time $\tau$ the atom will either leave the detection region or
stop scattering photons. Except for random background counts with
very low probability the detector sees no further light until the
next atom arrives (Fig.~\ref{fig:TIA}). Therefore for small atom
flux the photon count distribution should reflect both (i) the
instantaneous photon emission of the detected atom and its decay
at time scales $ \sim \tau$, and (ii) the statistical distribution
of the atoms at long time scales ($\gg\tau$).

To analyze these features, it is useful to measure the statistics
and time correlations of the photon counts. In the experiments
presented here we use thermal atoms in a multimode guide
(typically $>10^3$ transverse modes are occupied). Consequently,
the atoms exhibit Poissonian statistics as can be verified by a
time interval analysis of the atom arrival distribution
\cite{Kohl2007} described below .

In the case of a constant atom flux composed of uncorrelated
atoms, the probability of finding  $k$ consecutive bins that
contain no photons is given by
$P_{\mathrm{TI}}(k)=(1-p_{0})p_{0}^{k}$, where $p_0$ is the
probability for an empty bin. This means that $\log
P_{\mathrm{TI}}(k)$ is a linear function of $k$. In
Fig.~\ref{fig:TIA} we see that the time interval distribution is
composed of two exponential decays (red line). The steep slope for
short time intervals is determined by the instantaneous
fluorescence rate of individual atoms. For long time intervals the
slope is given by the atom arrival rate. Our measurements show
that in the latter region $\log p_{0}$ equals the mean rate of
atom detections, consistent with a Poissonian distribution of
arrival times. Any atom correlations present could be identified
from time interval analysis of single measurements
\cite{Kohl2007}.

An analysis of the photon noise shows that we can relate the
variance and mean of the photon counts directly to the average
number of photons detected from each atom \cite{man1965}. The
photon flux from a constant source during time intervals much
longer than the excited state life time $1/\Gamma$ is described by
a Poissonian probability distribution, where the mean photon
number and the variance are both equal to $\langle n\rangle$. If
the fluorescent photons come from a random flow of atoms described
by a statistical distribution $P_{\mathrm{atom}}(m)$ then $\langle
n\rangle$ is not constant in time. Mandel's formula must then be
used to describe the statistics of these photons \cite{tep2006}:
$P(n)=\sum_m P(n|m)P_{\mathrm{atom}}(m)$ where $P(n|m)$ is the
conditional probability of obtaining $n$ photons when the
observation region contains $m$ atoms. With the average number
$\alpha$ of photons detected per atom, defined by the relation
$\langle n\rangle=\alpha m$, the conditional photon distribution
is given by $ P(n|m)=\frac{(\alpha m)^n}{n!}\exp\left[-\alpha
m\right]. $

For the mean and the variance of the photon distribution one obtains the ratio
\begin{equation}\label{eq:poissonianAtomsNeglectBg}
\frac{\mathrm{var}(n_{\mathrm{photons}})}{\langle n_{\mathrm{photons}}\rangle}=1+\alpha\frac{\mathrm{var}(m_{\mathrm{atoms}})}{\langle m_{\mathrm{atoms}}\rangle}.
\end{equation}
When the atoms obey Poissonian statistics, as in our experiments,
Eq.(\ref{eq:poissonianAtomsNeglectBg}) reduces to
\mbox{$\mathrm{var}(n)/\langle n\rangle = 1 + \alpha$}. The mean
number of photons detected from each atom ($\alpha$) can be
directly retrieved from the ratio of variance to mean. If
additionally a Poissonian background $b$ is taken into account the
photon statistics can be expressed as
\begin{equation}\label{eq:poissonianAtomsAndBg}
\frac{\mathrm{var}(n_{\mathrm{photons}})}{\langle n_{\mathrm{photons}}\rangle}=\frac{(1+\alpha)\,\alpha\,\langle m_{\mathrm{atoms}}\rangle + \langle b \rangle}{\alpha\,\langle m_{\mathrm{atoms}}\rangle + \langle b \rangle}.
\end{equation}

The atom detection efficiency can be determined from $\alpha$. If
there is one atom in the detection region, it will generate at
least one photon count with a probability
\mbox{$P_{\mathrm{detection}}=1-\exp(-\alpha)$}.

From the photon statistics in the data shown in
Fig.~\ref{fig:varandmean}, one obtains $\alpha=1.08\pm0.01$ and
$P_{\mathrm{detection}}=66\%$ for 1~nW probe beam power, 3 MHz blue detuning, and
300~\textmu s integration time.

The total number of photons scattered by the atoms can be
independently obtained by measuring the ratio of the fluorescence
counts for F=2$\rightarrow$F'=1 and the F=2$\rightarrow$F'=3
transitions. On the F=2$\rightarrow$F'=1 transition an atom
scatters slightly more than one photon before being optically
pumped into the other hyperfine ground state, where it remains
dark. From the measured ratio we conclude that each atom scatters
$\sim{}120$ photons before it leaves the detector. These numbers
are in good agreement with $\alpha$=1.08 and the photon detection
efficiency \mbox{$p_\mathrm{det}=0.9$\%} given above.  In addition
the value of $\alpha$ was confirmed from independent global atom
number measurements using absorption imaging.

\begin{figure}[t!]
    \includegraphics[width=0.48\textwidth]{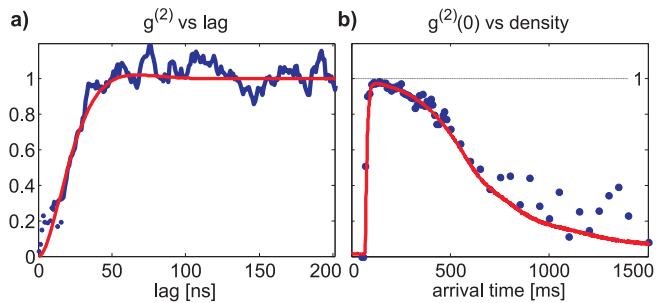}%
    \caption{(color online). second-order intensity correlation.
  \emph{a)} The second-order correlation function exhibits near perfect photon antibunching
    in the fluorescence emission of single atoms ($1.1$~nW excitation power, $s=3.5$).
    The red line is the theoretical model for the corresponding
    Rabi frequency $\Omega=2.3\Gamma$ \cite{Loudon}.
  \emph{b)} \gtwo{0} as function of arrival time.
    The graph covers atomic flux over three orders of magnitude
    (compare Fig.\ref{fig:varandmean}a).
    The red line is given by Eq.(\ref{eq:g2fit}).}
    \label{fig:g2}
\end{figure}

We can investigate if we really detect single atoms by looking at
the correlations in the detected fluorescence photons. Since a
single atom can emit only one photon at any given time one would
expect photon antibunching, characterized by $\gtwo{\delta{}t}<1$
\cite{Kimble77,Loudon}, where the second-order intensity
correlation function is given by
\begin{equation}
\label{eq:g2}
    \gtwo{\delta{}t}=
    \frac{\langle \hat{E}^-(t) \, \hat{E}^-(t+\delta{}t) \, \hat{E}^+(t+\delta{}t) \, \hat{E}^+(t) \rangle}
    {\langle \hat{E}^-(t) \, \hat{E}^+(t) \rangle^2}.
\end{equation}

Fig.~\ref{fig:g2}a shows a measurement of $\gtwo{\delta{}t}$ from
the cross-correlation of photon counts in two SPCMs arranged in a
Hanbury Brown-Twiss type setup as shown in Fig.\ref{fig:chippic}a.
The correlation function was reconstructed from the low density tail of the atom
distribution where the mean atomic distance is large enough to
guarantee the presence of at most one atom in the detection region
at any given time. An evaluation of the raw data results in a value
of $\gtwo{0}=0.05$. If we correct for coincidental background
counts,  \gtwo{0} is compatible
with zero. Thus we observe near-perfect photon antibunching in the
emission of single atoms passing the detector, a clear signature
of single atom detection. The single photon count rate is
approximately 3500 cps. The red line of Fig.~\ref{fig:g2}a is 
 the theoretically expected shape of \gtwo{\delta{}t} for
the employed excitation power according to \cite{Loudon}.

For a single-mode field with mean photon number $\langle n
\rangle$ the second-order correlation at lag $\delta{}t$=0 is limited by
\cite{Loudon}
\begin{equation}
    \gtwo{0}\geq{}1-\frac{1}{\langle n \rangle} \qquad  \forall \langle n \rangle \geq 1
\label{eq:g2n}
\end{equation}
while for $\langle n \rangle < 1$ the lower limit is 0. For a Fock
state with fixed photon number $n$ the inequality (\ref{eq:g2n})
becomes an equality and the minimal value of \gtwo{0} is reached.
While for classical light sources $1 \leq{} \gtwo{0} \leq{}
\infty$ holds the region $\gtwo{0} < 1$ is exclusively
nonclassical and can only reached by quantum emitters.

Since $\gtwo{\delta{}t}$ is evaluated when at least one photon
count has been recorded, the mean photon number $\langle n
\rangle$ has to be calculated under the condition $n \geq{} 1$.
This leads to \mbox{$\langle n \rangle = \alpha \langle N \rangle
/ (1 - \exp{(-\alpha \langle N \rangle)})$} with mean atom number
$\langle N \rangle$. Hence \gtwo{0} is limited by
\begin{equation}
    \gtwo{0}\geq{}1-\frac{1-\exp{(-\alpha \langle N \rangle)}}{\alpha \langle N \rangle}.
\label{eq:g2fit}
\end{equation}
As can be seen from Fig.~\ref{fig:g2}b the measured \gtwo{0}
follows the expected shape for the full atom pulse duration. With
this measurement we extend the original experimental investigation
\cite{Kimble77,Kimble78} of the influence of atomic density on the
second-order correlation function by almost three orders of
magnitude change in atomic density.

To conclude, we have built and evaluated an atom detector which is fully integrated on an atom chip, alignment free by fabrication, and mechanically very robust.  It is capable of detecting single atoms with 66\% efficiency and high signal to noise ratio, which allowed us to study the interplay between atom and photon statistics. 

Low noise, high efficiency and insensitivity to stray light is achieved using fiber optics to create very selective excitation of the atoms in a small, matched observation volume. The detection efficiency is currently limited by the numerical aperture of the multimode collection fiber. A straightforward substitution of the employed NA=0.275 fiber by a commercially available fiber with NA=0.53 increases the photon collection to $\alpha$ = 4.5 counts/atom and the single atom detection efficiency to 95\% at 50 kHz bandwidth. With these improvements, atom counting becomes feasible.

The high efficiency, signal-to-noise ratio and bandwidth of our integrated detector make it suitable for many physical systems where only a few photons can be scattered like in detecting single trapped cold molecules. With itÕs extremely low sensitivity to stray light, our detector is well suited for studies of correlated atomic systems and scalable quantum experiments on a single-atom or molecule level.

We thank T.~Fernholz, A.~Haase, and M.~Schwarz for help in the
early stages of the experiment and S.~Groth, I.~Bar-Joseph,
K.H.~Brenner and X.~Liu for fabrication of the chip and holding
structures. We gratefully acknowledge financial support from
Landesstiftung Baden-W{\"u}rtemberg, the and European Union
(SCALA, Atomchip, HIP), and the FWF (PLATON).


\begin{thebibliography}{34}
\expandafter\ifx\csname natexlab\endcsname\relax\def\natexlab#1{#1}\fi
\expandafter\ifx\csname bibnamefont\endcsname\relax
  \def\bibnamefont#1{#1}\fi
\expandafter\ifx\csname bibfnamefont\endcsname\relax
  \def\bibfnamefont#1{#1}\fi
\expandafter\ifx\csname citenamefont\endcsname\relax
  \def\citenamefont#1{#1}\fi
\expandafter\ifx\csname url\endcsname\relax
  \def\url#1{\texttt{#1}}\fi
\expandafter\ifx\csname urlprefix\endcsname\relax\def\urlprefix{URL }\fi
\providecommand{\bibinfo}[2]{#2}
\providecommand{\eprint}[2][]{\url{#2}}

\bibitem[{\citenamefont{Mandel and Wolf}(1995)}]{man1995}
\bibinfo{author}{\bibfnamefont{L.}~\bibnamefont{Mandel}} \bibnamefont{and}
  \bibinfo{author}{\bibfnamefont{E.}~\bibnamefont{Wolf}},
  \emph{\bibinfo{title}{{Optical Coherence and Quantum Optics}}}
  (\bibinfo{publisher}{Cambridge University Press}, \bibinfo{year}{1995}).

\bibitem[{\citenamefont{Bouwmeester et~al.}(2001)\citenamefont{Bouwmeester,
  Ekert, and Zeilinger}}]{Bow00}
\bibinfo{author}{\bibfnamefont{D.}~\bibnamefont{Bouwmeester}},
  \bibinfo{author}{\bibfnamefont{A.}~\bibnamefont{Ekert}}, \bibnamefont{and}
  \bibinfo{author}{\bibfnamefont{A.}~\bibnamefont{Zeilinger}},
  \emph{\bibinfo{title}{The Physics of Quantum Information: Quantum
  Cryptography, Quantum Teleportation, Quantum Computation}}
  (\bibinfo{publisher}{Springer Verlag}, \bibinfo{address}{Heidelberg},
  \bibinfo{year}{2001}).

\bibitem[{\citenamefont{Mabuchi et~al.}(1996)\citenamefont{Mabuchi, Turchette,
  Chapman, and Kimble}}]{mab1996}
\bibinfo{author}{\bibfnamefont{H.}~\bibnamefont{Mabuchi}},
  \bibinfo{author}{\bibfnamefont{Q.}~\bibnamefont{Turchette}},
  \bibinfo{author}{\bibfnamefont{M.}~\bibnamefont{Chapman}}, \bibnamefont{and}
  \bibinfo{author}{\bibfnamefont{H.}~\bibnamefont{Kimble}},
  \bibinfo{journal}{Opt. Lett} \textbf{\bibinfo{volume}{21}},
  \bibinfo{pages}{1393} (\bibinfo{year}{1996}).

\bibitem[{\citenamefont{M{\"u}nstermann
  et~al.}(1999)\citenamefont{M{\"u}nstermann, Fischer, Maunz, Pinkse, and
  Rempe}}]{mun1999}
\bibinfo{author}{\bibfnamefont{P.}~\bibnamefont{M{\"u}nstermann}},
  \bibinfo{author}{\bibfnamefont{T.}~\bibnamefont{Fischer}},
  \bibinfo{author}{\bibfnamefont{P.}~\bibnamefont{Maunz}},
  \bibinfo{author}{\bibfnamefont{P.}~\bibnamefont{Pinkse}}, \bibnamefont{and}
  \bibinfo{author}{\bibfnamefont{G.}~\bibnamefont{Rempe}},
  \bibinfo{journal}{Phys. Rev. Lett.} \textbf{\bibinfo{volume}{82}},
  \bibinfo{pages}{3791} (\bibinfo{year}{1999}).

\bibitem[{\citenamefont{Hood et~al.}(2000)\citenamefont{Hood, Lynn, Doherty,
  Parkins, and Kimble}}]{hoo2000}
\bibinfo{author}{\bibfnamefont{C.}~\bibnamefont{Hood}},
  \bibinfo{author}{\bibfnamefont{T.}~\bibnamefont{Lynn}},
  \bibinfo{author}{\bibfnamefont{A.}~\bibnamefont{Doherty}},
  \bibinfo{author}{\bibfnamefont{A.}~\bibnamefont{Parkins}}, \bibnamefont{and}
  \bibinfo{author}{\bibfnamefont{H.}~\bibnamefont{Kimble}},
  \bibinfo{journal}{Science} \textbf{\bibinfo{volume}{287}},
  \bibinfo{pages}{1447} (\bibinfo{year}{2000}).

\bibitem[{\citenamefont{{\"O}ttl et~al.}(2005)\citenamefont{{\"O}ttl, Ritter,
  K{\"o}hl, and Esslinger}}]{ott2005}
\bibinfo{author}{\bibfnamefont{A.}~\bibnamefont{{\"O}ttl}},
  \bibinfo{author}{\bibfnamefont{S.}~\bibnamefont{Ritter}},
  \bibinfo{author}{\bibfnamefont{M.}~\bibnamefont{K{\"o}hl}}, \bibnamefont{and}
  \bibinfo{author}{\bibfnamefont{T.}~\bibnamefont{Esslinger}},
  \bibinfo{journal}{Phys. Rev. Lett.} \textbf{\bibinfo{volume}{95}},
  \bibinfo{pages}{090404} (\bibinfo{year}{2005}).

\bibitem[{\citenamefont{Haase et~al.}(2006)\citenamefont{Haase, Hessmo, and
  Schmiedmayer}}]{Haa06}
\bibinfo{author}{\bibfnamefont{A.}~\bibnamefont{Haase}},
  \bibinfo{author}{\bibfnamefont{B.}~\bibnamefont{Hessmo}}, \bibnamefont{and}
  \bibinfo{author}{\bibfnamefont{J.}~\bibnamefont{Schmiedmayer}},
  \bibinfo{journal}{Opt. Lett.} \textbf{\bibinfo{volume}{31}},
  \bibinfo{pages}{268} (\bibinfo{year}{2006}).

\bibitem[{\citenamefont{Aoki et~al.}(2006)\citenamefont{Aoki, Dayan, Wilcut,
  Bowen, Parkins, Kippenberg, Valhala, and Kimble}}]{aok2006}
\bibinfo{author}{\bibfnamefont{T.}~\bibnamefont{Aoki}},\emph{ et al.,}
  \bibinfo{journal}{Nature(London)} \textbf{\bibinfo{volume}{443}},
  \bibinfo{pages}{671} (\bibinfo{year}{2006}).

\bibitem[{\citenamefont{Teper et~al.}(2006)\citenamefont{Teper, Lin, and
  Vuleti{\'c}}}]{tep2006}
\bibinfo{author}{\bibfnamefont{I.}~\bibnamefont{Teper}},
  \bibinfo{author}{\bibfnamefont{Y.}~\bibnamefont{Lin}}, \bibnamefont{and}
  \bibinfo{author}{\bibfnamefont{V.}~\bibnamefont{Vuleti{\'c}}},
  \bibinfo{journal}{Phys. Rev. Lett.} \textbf{\bibinfo{volume}{97}},
  \bibinfo{pages}{023002} (\bibinfo{year}{2006}).

\bibitem[{\citenamefont{Colombe et~al.}(2007)\citenamefont{Colombe, Steinmetz,
  Dubois, Linke, Hunger, and Reichel}}]{col2007}
\bibinfo{author}{\bibfnamefont{Y.}~\bibnamefont{Colombe}},\emph{ et al.,}
  \bibinfo{journal}{Nature(London)} \textbf{\bibinfo{volume}{450}},
  \bibinfo{pages}{272} (\bibinfo{year}{2007}).

\bibitem[{\citenamefont{Trupke et~al.}(2007)\citenamefont{Trupke, Goldwin,
  Darqui\'{e}, Dutier, Eriksson, Ashmore, and Hinds}}]{tru2007}
\bibinfo{author}{\bibfnamefont{M.}~\bibnamefont{Trupke}},\emph{ et al.,}
  \bibinfo{journal}{Phys. Rev. Lett.} \textbf{\bibinfo{volume}{99}},
  \bibinfo{pages}{063601} (\bibinfo{year}{2007}).

\bibitem[{\citenamefont{van Enk and Kimble}(2001)}]{vEnk01}
\bibinfo{author}{\bibfnamefont{S.~J.} \bibnamefont{van Enk}} \bibnamefont{and}
  \bibinfo{author}{\bibfnamefont{H.~J.} \bibnamefont{Kimble}},
  \bibinfo{journal}{Phys. Rev. A} \textbf{\bibinfo{volume}{63}},
  \bibinfo{pages}{023809} (\bibinfo{year}{2001}).

\bibitem[{\citenamefont{Quinto-Su et~al.}(2004)\citenamefont{Quinto-Su,
  Tscherneck, Holmes, and Bigelow}}]{quintosu2004}
\bibinfo{author}{\bibfnamefont{P.}~\bibnamefont{Quinto-Su}},
  \bibinfo{author}{\bibfnamefont{M.}~\bibnamefont{Tscherneck}},
  \bibinfo{author}{\bibfnamefont{M.}~\bibnamefont{Holmes}}, \bibnamefont{and}
  \bibinfo{author}{\bibfnamefont{N.}~\bibnamefont{Bigelow}},
  \bibinfo{journal}{Opt. Expr.} \textbf{\bibinfo{volume}{12}},
  \bibinfo{pages}{5098} (\bibinfo{year}{2004}).

\bibitem[{\citenamefont{Tey et~al.}(2008)\citenamefont{Tey, Chen, Aljunid,
  Chng, Huber, Maslennikov, and Kurtsiefer}}]{tey2008}
\bibinfo{author}{\bibfnamefont{M.}~\bibnamefont{Tey}},\emph{ et al.,}
  \bibinfo{journal}{Arxiv preprint arXiv:0802.3005}  (\bibinfo{year}{2008}).

\bibitem[{\citenamefont{Kuhr et~al.}(2001)\citenamefont{Kuhr, Alt, Schrader,
  Muller, Gomer, and Meschede}}]{kuh2001}
\bibinfo{author}{\bibfnamefont{S.}~\bibnamefont{Kuhr}},\emph{ et al.,}
  \bibinfo{journal}{Science} \textbf{\bibinfo{volume}{293}},
  \bibinfo{pages}{278} (\bibinfo{year}{2001}).

\bibitem[{\citenamefont{Rowe et~al.}(2001)\citenamefont{Rowe, Kielpinski,
  Meyer, Sackett, Itano, Monroe, and Wineland}}]{row2001}
\bibinfo{author}{\bibfnamefont{M.}~\bibnamefont{Rowe}},\emph{ et al.,}
  \bibinfo{journal}{Nature} \textbf{\bibinfo{volume}{409}},
  \bibinfo{pages}{791} (\bibinfo{year}{2001}).

\bibitem[{\citenamefont{Leibfried et~al.}(2003)\citenamefont{Leibfried, Blatt,
  Monroe, and Wineland}}]{lei2003}
\bibinfo{author}{\bibfnamefont{D.}~\bibnamefont{Leibfried}},
  \bibinfo{author}{\bibfnamefont{R.}~\bibnamefont{Blatt}},
  \bibinfo{author}{\bibfnamefont{C.}~\bibnamefont{Monroe}}, \bibnamefont{and}
  \bibinfo{author}{\bibfnamefont{D.}~\bibnamefont{Wineland}},
  \bibinfo{journal}{Rev. Mod. Phys.} \textbf{\bibinfo{volume}{75}},
  \bibinfo{pages}{281} (\bibinfo{year}{2003}).

\bibitem[{\citenamefont{Darquie et~al.}(2005)\citenamefont{Darquie, Jones,
  Dingjan, Beugnon, Bergamini, Sortais, Messin, Browaeys, and
  Grangier}}]{dar2005}
\bibinfo{author}{\bibfnamefont{B.}~\bibnamefont{Darquie}},\emph{ et al.,}
  \bibinfo{journal}{Science} \textbf{\bibinfo{volume}{309}},
  \bibinfo{pages}{454} (\bibinfo{year}{2005}).

\bibitem[{\citenamefont{Volz et~al.}(2006)\citenamefont{Volz, Weber, Schlenk,
  Rosenfeld, Vrana, Saucke, Kurtsiefer, and Weinfurter}}]{vol2006}
\bibinfo{author}{\bibfnamefont{J.}~\bibnamefont{Volz}},\emph{ et al.,}
  \bibinfo{journal}{Phys. Rev. Lett.} \textbf{\bibinfo{volume}{96}},
  \bibinfo{pages}{030404} (\bibinfo{year}{2006}).

\bibitem[{\citenamefont{Weber et~al.}(2006)\citenamefont{Weber, Volz, Saucke,
  Kurtsiefer, and Weinfurter}}]{web2006}
\bibinfo{author}{\bibfnamefont{M.}~\bibnamefont{Weber}},
  \bibinfo{author}{\bibfnamefont{J.}~\bibnamefont{Volz}},
  \bibinfo{author}{\bibfnamefont{K.}~\bibnamefont{Saucke}},
  \bibinfo{author}{\bibfnamefont{C.}~\bibnamefont{Kurtsiefer}},
  \bibnamefont{and}
  \bibinfo{author}{\bibfnamefont{H.}~\bibnamefont{Weinfurter}},
  \bibinfo{journal}{Phys. Rev.} \textbf{\bibinfo{volume}{A73}},
  \bibinfo{pages}{043406} (\bibinfo{year}{2006}).

\bibitem[{\citenamefont{Sortais et~al.}(2007)}]{sor2007}
\bibinfo{author}{\bibfnamefont{Y.~R.~P.} \bibnamefont{Sortais}}
  \bibnamefont{\emph{ et al.}}, \bibinfo{journal}{Phys. Rev.}
  \textbf{\bibinfo{volume}{A75}}, \bibinfo{pages}{013406}
  (\bibinfo{year}{2007}).

\bibitem[{\citenamefont{Bondo et~al.}(2006)\citenamefont{Bondo, Hennrich,
  Legero, Rempe, and Kuhn}}]{bon2006}
\bibinfo{author}{\bibfnamefont{T.}~\bibnamefont{Bondo}},
  \bibinfo{author}{\bibfnamefont{M.}~\bibnamefont{Hennrich}},
  \bibinfo{author}{\bibfnamefont{T.}~\bibnamefont{Legero}},
  \bibinfo{author}{\bibfnamefont{G.}~\bibnamefont{Rempe}}, \bibnamefont{and}
  \bibinfo{author}{\bibfnamefont{A.}~\bibnamefont{Kuhn}},
  \bibinfo{journal}{Opt. Comm.} \textbf{\bibinfo{volume}{264}},
  \bibinfo{pages}{271} (\bibinfo{year}{2006}).

\bibitem[{\citenamefont{Folman et~al.}(2002)\citenamefont{Folman, Kr\"uger,
  Schmiedmayer, Denschlag, and Henkel}}]{Fol02}
\bibinfo{author}{\bibfnamefont{R.}~\bibnamefont{Folman}},\emph{ et al.,}
  \bibinfo{journal}{Adv. At. Mol. Phys.} \textbf{\bibinfo{volume}{48}},
  \bibinfo{pages}{263} (\bibinfo{year}{2002}).

\bibitem[{\citenamefont{Fort\'{a}gh and Zimmermann}(2007)}]{Zimmermann2007}
\bibinfo{author}{\bibfnamefont{J.}~\bibnamefont{Fort\'{a}gh}} \bibnamefont{and}
  \bibinfo{author}{\bibfnamefont{C.}~\bibnamefont{Zimmermann}},
  \bibinfo{journal}{Rev. Mod. Phys.} \textbf{\bibinfo{volume}{79}},
  \bibinfo{pages}{235} (\bibinfo{year}{2007}).

\bibitem[{\citenamefont{Abraham and Cornell}(1998)}]{abr1998}
\bibinfo{author}{\bibfnamefont{E.}~\bibnamefont{Abraham}} \bibnamefont{and}
  \bibinfo{author}{\bibfnamefont{E.}~\bibnamefont{Cornell}},
  \bibinfo{journal}{Appl. Opt} \textbf{\bibinfo{volume}{37}},
  \bibinfo{pages}{1762} (\bibinfo{year}{1998}).

\bibitem[{\citenamefont{Groth et~al.}(2004)\citenamefont{Groth, Kruger,
  Wildermuth, Folman, Fernholz, Schmiedmayer, Mahalu, and
  Bar-Joseph}}]{groth2004}
\bibinfo{author}{\bibfnamefont{S.}~\bibnamefont{Groth}}, \emph{et al.,}
  \bibinfo{journal}{Appl. Phys. Lett.} \textbf{\bibinfo{volume}{85}},
  \bibinfo{pages}{2980} (\bibinfo{year}{2004}).

\bibitem[{\citenamefont{Liu et~al.}(2005)}]{liu2005}
\bibinfo{author}{\bibfnamefont{X.}~\bibnamefont{Liu}}, \emph{et al.,}
  \bibinfo{journal}{Appl. Opt} \textbf{\bibinfo{volume}{44}},
  \bibinfo{pages}{6857} (\bibinfo{year}{2005}).

\bibitem[{\citenamefont{Wilzbach et~al.}(2006)}]{wil2006}
\bibinfo{author}{\bibfnamefont{M.}~\bibnamefont{Wilzbach}}, \emph{et al.,}\bibinfo{journal}{Fortschritte der Physik}
  \textbf{\bibinfo{volume}{54}}, \bibinfo{pages}{746} (\bibinfo{year}{2006}).

\bibitem[{\citenamefont{Wildermuth et~al.}(2004)}]{wil2004}
\bibinfo{author}{\bibfnamefont{S.}~\bibnamefont{Wildermuth}}
 \emph{et al.,} \bibinfo{journal}{Phys. Rev.}
  \textbf{\bibinfo{volume}{A69}}, \bibinfo{pages}{030901(R)}
  (\bibinfo{year}{2004}).

\bibitem[{\citenamefont{K{\"o}hl et~al.}(2007)\citenamefont{K{\"o}hl, {\"O}ttl,
  Ritter, Donner, Bourdel, and Esslinger}}]{Kohl2007}
\bibinfo{author}{\bibfnamefont{M.}~\bibnamefont{K{\"o}hl}}, \emph{et al.,}
  \bibinfo{journal}{Appl. Phys. B} \textbf{\bibinfo{volume}{86}},
  \bibinfo{pages}{391} (\bibinfo{year}{2007}).

\bibitem[{\citenamefont{Mandel and Wolf}(1965)}]{man1965}
\bibinfo{author}{\bibfnamefont{L.}~\bibnamefont{Mandel}} \bibnamefont{and}
  \bibinfo{author}{\bibfnamefont{E.}~\bibnamefont{Wolf}},
  \bibinfo{journal}{Rev. Mod. Phys.} \textbf{\bibinfo{volume}{37}},
  \bibinfo{pages}{231} (\bibinfo{year}{1965}).

\bibitem[{\citenamefont{Loudon}(2000)}]{Loudon}
\bibinfo{author}{\bibfnamefont{R.}~\bibnamefont{Loudon}},
  \emph{\bibinfo{title}{{T}he {Q}uantum {T}heory of {L}ight}}
  (\bibinfo{publisher}{Oxford Science}, \bibinfo{year}{2000}).

\bibitem[{\citenamefont{Kimble et~al.}(1977)\citenamefont{Kimble, Dagenais, and
  Mandel}}]{Kimble77}
\bibinfo{author}{\bibfnamefont{H.~J.} \bibnamefont{Kimble}},
  \bibinfo{author}{\bibfnamefont{M.}~\bibnamefont{Dagenais}}, \bibnamefont{and}
  \bibinfo{author}{\bibfnamefont{L.}~\bibnamefont{Mandel}},
  \bibinfo{journal}{Phys. Rev. Lett.} \textbf{\bibinfo{volume}{39}},
  \bibinfo{pages}{691} (\bibinfo{year}{1977}).

\bibitem[{\citenamefont{Kimble et~al.}(1978)\citenamefont{Kimble, Dagenais, and
  Mandel}}]{Kimble78}
\bibinfo{author}{\bibfnamefont{H.~J.} \bibnamefont{Kimble}},
  \bibinfo{author}{\bibfnamefont{M.}~\bibnamefont{Dagenais}}, \bibnamefont{and}
  \bibinfo{author}{\bibfnamefont{L.}~\bibnamefont{Mandel}},
  \bibinfo{journal}{Phys. Rev. A} \textbf{\bibinfo{volume}{18}},
  \bibinfo{pages}{201} (\bibinfo{year}{1978}).

\end{thebibliography}

\end{document}